# A Method for Fabricating CMOS Back-End-of-Line-Compatible Solid-State Nanopore Devices

Mohamed Yassine Bouhamidi[2,#], Chunhui Dai[1,#], Michel Stephan[2], Joyeeta Nag[1], Justin Kinney[1], Lei Wan[1], Matthew Waugh[2,3], Kyle Briggs[2,3], Jordan Katine[1], Vincent Tabard-Cossa[2,3,†], Daniel Bedau[1,*]

1. Western Digital Research, San Jose, California, United States
2. Department of Physics, University of Ottawa, Ottawa, Ontario, Canada
3. Northern Nanopore Instruments (Canada).

\# Equal contribution

[†]Corresponding author: tcossa@uOttawa.ca

*Corresponding author: daniel.bedau@wdc.com

# Abstract:

Solid-state nanopores, nm-sized holes in thin, freestanding membranes, are powerful single-molecule sensors capable of interrogating a wide range of target analytes, from small molecules to large polymers. Interestingly, due to their high spatial resolution, nanopores can also identify tags on long polymers, making them an attractive option as the reading element for molecular information storage strategies. To fully leverage the compact and robust nature of solid-state nanopores, however, they will need to be packaged in a highly parallelized manner with on-chip electronic signal processing capabilities to rapidly and accurately handle the data generated. Additionally, the membrane itself must have specific physical, chemical, and electrical properties to ensure sufficient signal-to-noise ratios are achieved, with the traditional membrane material being $SiN_X$. Unfortunately, the typical method of deposition, low-pressure vapour deposition, requires temperatures beyond the thermal budget of CMOS back-end-of-line integration processes, limiting the potential to generate an on-chip solution. To this end, we explore various lower-temperature deposition techniques that are BEOL-compatible to generate SiNx membranes for solid-state nanopore use, and successfully demonstrate the ability for these alternative methods to generate low-noise nanopores that are capable of performing single-molecule experiments.

# Section 1: Introduction

As the total volume of global data continues to soar, conventional storage methods such as tape, hard disk drives (HDD), and solid-state drives (SSD) are struggling to keep pace [1]. This burgeoning demand has spurred considerable interest in exploring alternative data storage solutions. One such solution, molecular information storage (MIST), has emerged as a promising candidate, particularly for archival purposes, owing to its exceptional data density [2, 3], long-term stability [4], and remarkable energy efficiency [5]. However, in order for the promise of MIST to be realized, advancements in both the data writing element (i.e., molecular synthesis) and the data reading element (i.e. reading the sequence of data-encoded molecules) are required [2]. Among the various molecular sequencing techniques available, nanopores represent a paradigm shift due to their potential for long-read and high-throughput bit detection [6-11] and in particular, solid-state nanopores (ssNPs) offer additional advantages, including robustness and potential compatibility with large-scale electronics integration techniques [8-11] when compared to their biological counterparts.

To date, however, integrating ssNPs into large-scale electronics has not yet been realized. One major reason for this is the manner through which the membrane film is deposited. Typically, ssNPs are fabricated in $SiN_X$ membranes which are deposited on a substrate (typically Si) through low-pressure chemical vapour deposition (LPCVD). This technique is based on the thermal decomposition of the gas-phase reactants (silane, $SiH_4$ and $NH_3$) on the wafer surface [12]. While the LPCVD process yields a slow deposition rate, it produces high-performance films with uniformity and low hydrogen content [12] and provides a reliable thin film (~10-20 nm) in which nanopores can be fabricated. However, in order for this reaction to proceed as described, the deposition process itself needs to be performed at 700°C-1000°C [13]. This presents a problem for complementary metal-oxide semiconductor (CMOS) back-end-of-line (BEOL) integration as a low thermal budget (<400°C) is often required to protect the low-k dielectric and copper interconnects of the devices from undergoing dielectric breakdown and negatively impacting reliability. Therefore, for ssNP to leverage the signal processing power, system-level miniaturization, and low-cost mass production capabilities afforded by CMOS and BEOL integration, an alternative membrane deposition method must be identified.

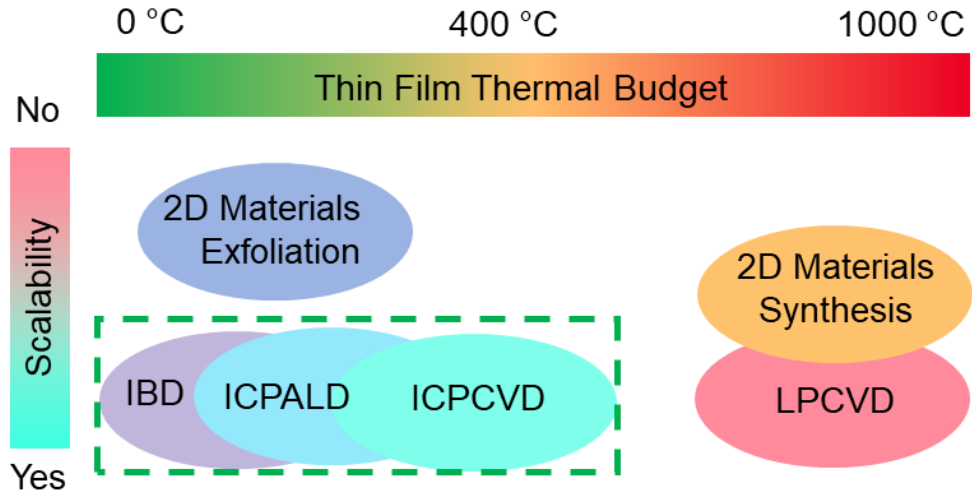

Figure 1. Assessment of the scalability and typical process temperatures of various film generation techniques. In particular, we highlight low-pressure vapour deposition (LPCVD), 2D material exfoliation, 2D material synthesis, ion beam deposition (IBD), inductively coupled plasma chemical vapor deposition (ICPCVD), and plasma enhanced chemical vapor deposition (PECVD), specifically inductively coupled plasma enhanced ALD (ICPALD). The dotted rectangle highlights a regime that is both scalable and at a temperature low enough to be compatible with complementary metal-oxide semiconductor back-end-of-line integration.

Fortunately, there are a number of membrane deposition approaches that could provide suitable membranes for ssNP devices while also aligning with the thermal budget for BEOL integration (Figure 1). We identify three potentially suitable methods: i) ion beam deposition (IBD), which operates at room temperature, ii) inductively coupled plasma chemical vapor deposition (ICPCVD), which is a special form of plasma enhanced chemical vapor deposition (PECVD) and operates at approximately 200-400°C [14], and iii) inductively coupled plasma enhanced atomic layer deposition (ICPALD), which is a type of plasma enhanced ALD (PEALD), which operates at approximately 200°C and achieves higher plasma densities and better control than other types of PEALD.

However, as both ICPCVD and PEALD use plasma ionization to enable deposition at a much lower temperature than the temperatures at which the precursors (e.g. $SiH_4$, $NH_3$, $N_2$, organic precursor such as bis(tert-butylamino)silane (BTBAS)) would naturally react, some of the precursors can remain unreacted or incompletely-reacted, leading to an elevated presence of carbon and hydrogen in the films [15,16]. This may cause enhanced porosity in such films, rendering them less preferable for downstream processes for nanopore fabrication. In contrast, IBD deposits $SiN_X$ films using nitrogen plasma to react with Si ions ejected from a pure Si target [17]. IBD is a physical vapor deposition (PVD) technique that produces thin films using inert gas plasma to dislodge film-forming particles from a desired target material onto a substrate with many desirable properties, for instance, improved adhesion, dense film structure, fewer defects, higher purity, and better controlled composition compared to films grown by other PVD techniques. This is related to various distinguishing features of IBD – (1) the working pressure is low, typically less than $10^{-1}$ Torr; (2) the ion generation and acceleration (ion beam source), generation of film-forming particles (target), and thin film growth (substrate) are spatially separated, thus geometrical parameters (ion incidence angle and emission angle) in addition to ion beam parameters (ion species and energy) can be varied; (3) another big difference between conventional ion beam deposition and magnetron sputtering is that for IBD there is no plasma between the substrate and the target, therefore IBD allows sensitive substrates to be deposited on and also the sputter gas inclusion in the deposit is minimal, increasing density of films [18]. In contrast to conventional sputtering, in IBD there is no bias between the substrate and the target, meaning both conducting as well as non-conducting targets and substrates can be used, avoiding the undesirable arcing during deposition altogether (arcing during depositions can cause significantly lower quality films). In comparison to CVD processes, in such a purely physical deposition process no organic precursor is involved and thus the presence of hydrogen and carbon are minimal in the film. As a result, we deemed IBD as the optimal process compatible with the needs of both solid-state nanopore use as well as BEOL.

In addition to the type of deposition, we also identified thermal annealing as a potential method for improving the membrane for use with ssNPs. By subjecting the membrane to elevated temperatures and slowly allowing them to cool, we sought to reduce the stress within the membrane itself, which may aid in the formation of more stable, lower noise ssNPs.

We therefore present an analysis of the $SiN_X$ film resulting from ion beam deposition on Si as well as a full characterization of its suitability for ssNP work including nanopore fabrication via Controlled Breakdown (CBD), electrical noise response, size stability, and the ability to detect single molecules for both IBD and annealed IBD membranes. Our findings pave the way for the continued development of a CMOS-integrated ssNP system for high-throughput sequencing, particularly for MIST applications.

# Section 2: Fabrication of Nanopore Devices

## A. Ion Beam Deposition of $SiN_x$

The IBD $SiN_x$ film was prepared by first inserting the substrate, 100 mm intrinsic, undoped high resistivity (>10 000 Ω-cm) silicon wafers with a thickness of 300 ± 20 µm, into an ion beam deposition system (Veeco) with its vacuum chamber pumped down to $10^{-8}$ Torr. A Si target (99.999% purity) was used as a source of ions for the deposition. Si atoms were liberated from the target by Ar ions and subsequently reacted with nitrogen ions (N2 gas flow was fixed at 35sccm). The SiNx molecules then settled onto the substrate, forming a thin $SiN_X$ film, with the thickness being determined by the duration of the deposition. For our purposes, 100 nm $SiN_X$ was deposited for the membrane devices. The ion beam was operated at 550V with a beam current of 200 mA.

To anneal the $SiN_x$ thin film, the wafer was loaded into a rapid thermal annealing (RTA) system where the chamber was pumped down to $10^{-5}$ Torr and then purged with Ar gas with a flow rate of 40 sccm to maintain 5 Torr chamber pressure. The system temperature was then quickly ramped up (at 1 °C/s) to 800 °C and maintained for 240 s. After annealing, the system was cooled to room temperature with a continuous Ar gas flow before the sample was removed from the chamber.

## B. Membrane Fabrication

Once the $SiN_X$ deposition was complete, membrane devices were fabricated by standard photolithography and microfabrication procedures using two photomasks (soda-lime glass with chrome coating). The first photomask, used on the cleanest side of the wafer (referred to as *membrane side*), featured 5 µm circular patterns that were designed to be centered within the free-standing membranes that would be created. These circular features defined the region where the membrane would be locally thinned from 100 nm down to 20 nm to promote nanopore formation in that location. Additionally, the *membrane side* photomask contained the grid patterns necessary for dicing the wafer into 5×5 mm square chips. The second photomask, which is used on the other side of the wafer (referred to as *etched pit side*), was designed to pattern specifically dimensioned square windows which would be centralized to the previously patterned local thinning circles. The square windows represent the region etched by potassium hydroxide (KOH) to produce the 40×40 µm free standing membranes. These two photomasks were each designed with the same coarse alignment marks which were used to align the wafer to the photomask to maximize the number of patterns that can fit on the surface of the wafer. Additionally, the photomasks featured different but complementary precision alignment marks that enabled double-sided alignment of the wafer and ensured the patterns on each side of its sides were centralized one to another.

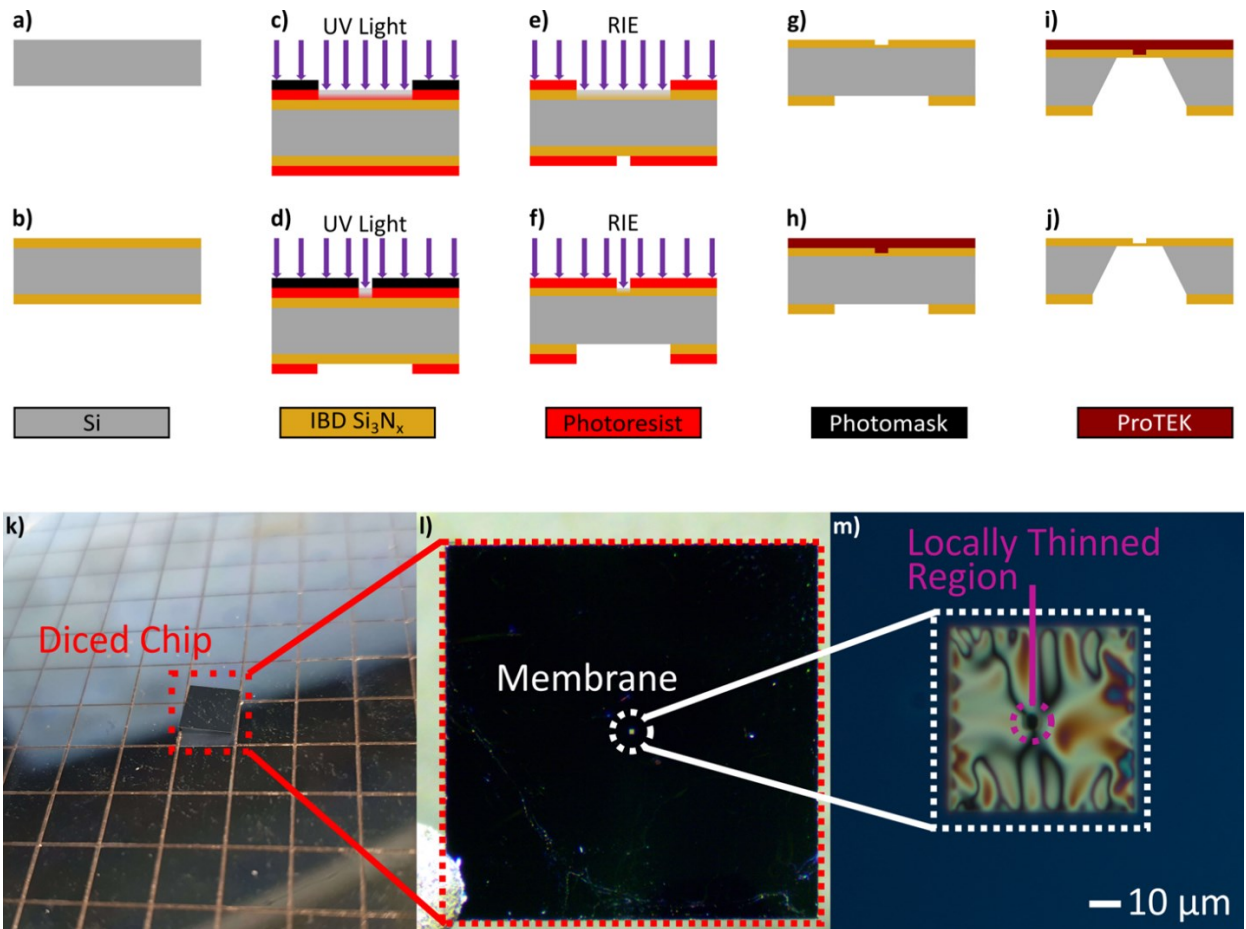

Figure 2. Overview of the IBD-based nanofabrication process to generate a $SiN_X$ membrane (a-j) and associated resultant images (k-m). a) Bare double-side polished silicon wafer with crystal orientation <100>. b) Ion beam deposition (IBD) of $SiN_X$ on both sides of the wafer generates a film of ~100 nm. C) Alignment of the etched pit side of the wafer with its appropriate photomask and patterning, through UV photolithography, of the KOH etch windows. d) Alignment of the membrane side of the wafer with its appropriate photomask and patterning, through UV photolithography, of the local thinning circular regions that will be positioned in the center of each membrane. e) Reactive ion etching (RIE) of the entire thickness of the SiNx exposed features on the etched pit side of the wafer to define square membrane opening for later KOH etching step. f) RIE of a partial thickness of the SiNx (until an approximate thickness of 10-20 nm remains) exposed features on the membrane side of the wafer to locally thin a portion of the membrane

which will enforce nanopore fabrication to this region. g) Photoresist stripping. h) Spin coating of a protective polymer layer, ProTEK, on the membrane side of the wafer. i) Hot KOH etching of the exposed silicon on the etched pit side of the wafer to form and release free stranding thin films. j) Stripping of the protective polymer layer after the wafer is diced into individual 5×5mm chips. k) Picture depicting a diced wafer with a selected chip. l) Close-up image of the chip and the membrane it contains. m) Microscopy image of the IBD SiNx membrane on the chosen chip.

The fabrication of the membrane devices for nanopore use was performed as outlined in Figure 2 a-j [19], starting by spin-coating a thin (~650 nm) layer of positive photoresist (*Microposit S1805*) on each side of the wafer, which was then soft baked on a hot plate (115°C for 60s). Next, with a mask aligner, rough alignment of the *membrane side* of the wafer to the first photomask was conducted using the patternable bottom alignment marks and the wafer primary flat. Once aligned, the wafer was exposed to UV light that rendered the photoresist (PR) soluble where patterns were desired. The photolithographed wafer was then put into a developer solution (*Microposit MF-321*) that dissolved the exposed PR and visually defined the patterned features and the precision alignment marks (Figure 2c). At that point, the wafer was turned over and the *etched pit side* and loaded in the mask-aligner where the second photomask would rest on top of it. The initial alignment was performed as before, using the bottom alignment marks and wafer primary flat. Once rough alignment was achieved, greater precision was reached by using an infrared lamp and aligning the patterned alignment marks on the *membrane side* of the wafer with the complementary alignment marks on the second photomask. The wafer was then exposed and developed as previously done (Figure 2d). Reactive ion etching (RIE) of the *etched pit side* of the wafer was then executed following a standard recipe (5 sccm of $O_2$, 25 sccm of $CF_4$, 100 W, 8 Pa, 45 s) to ablate the exposed $SiN_x$ patterns through their entire thickness, until the silicon substrate was reached. This step produced exposed silicon windows, which were regions where KOH would etch the wafer to release free-standing membranes in the upcoming steps (Figure

2e). Next, with the objective of retaining a very thin $SiN_X$ layer for the future thin film structure, a low-power and slower RIE step (10 sccm of $CF_4$, 25 W, 8 Pa, 4.5 to 6 min) was performed on the *membrane side* of the wafer to obtain a locallythinned region that was below 20 nm in thickness on average (Figure 2f). Remaining photoresist was then stripped of both sides of the wafer by sequentially submerging it into separate solutions of acetone, isopropyl alcohol, and deionized water (Figure 2g). A primer and protective polymer (*Brewer Science ProTEK B3-Primer* and *B3-25*) were spin coated and baked (205°C for 60s for the primer and 120°C for 120s followed by 205°C for 60s for the protective polymer) on the membrane side of the device to prevent formation of pinholes during the following wet etch (Figure 2h). To release the $SiN_x$ thin film structures, the wafer underwent a wet etch step by being submerged for 4 hours in a 75 °C KOH solution with a weight-by-weight concentration of 30%. In this solution, any exposed silicon surface would be attacked the KOH and follow an anisotropic etch with a characteristic angle of 54.7° between the wafer crystalline plane <100> and <111> [20] (Figure 2i). The wafer was then diced in individual 5×5 mm chips as shown in Figure 2k. The precisely diced chips were then subjected to a cleaning step involving a sequential Piranha at 90 ºC for 1 hour and an RCA1 procedure [20] to fully remove the *ProTEK* protective film and any surface contamination (Figure 2j). Finally, the cleaned chips were stored in 0.6 mL microcentrifuge tubes filled with a 1:1 solution of ethanol and DI water until they were used.

### C. Nanopore Fabrication and Use

Nanopores were fabricated in-situ by the controlled breakdown (CBD) method [21]. To do this, the chips were first removed from their storage buffer, carefully dried by using a weak vacuum or a soft stream of N$_2$ gas and mounted into 3D printed flow cells (Northern Nanopore Instruments). Isopropyl alcohol (IPA) was then flushed through each side of the flow cell to ensure proper wetting of the membrane, followed by DI water and then a solution of 1M KCl + 10 mM HEPES pH 8. Electrical contact was then made via a pair of Ag/AgCl electrodes of 1 mm in diameter that were immersed in the salt solution. The electrodes were held in place by caps that seal the fluidic chambers to reduce evaporation of the solution. The flow cell was finally placed into the Spark-E2 unit (Northern Nanopores Instruments) for automated nanopore fabrication, and the associated software commanded to fabricate a nanopore of the size desired using the Standard Protocol. Once the nanopore was fabricated via controlled breakdown, 3.6 M LiCl + 10 mM HEPES pH 8 was flushed into the flow cell and the software commanded to enter the conditioning stage where it enlarged the nanopore to the desired size using series of 5 voltage pulses at 40% of the breakdown voltage for 1 s each. Once completed, a power spectral density (PSD) at 200 mV was performed to characterize the electrical noise of the nanopore using a low-noise current amplifier (Axopatch 200B, Molecular Devices). For nanopores with sufficiently low 1/$f$ noise, defined here for most pores as <10 pA$^2$/Hz at 1 Hz on the PSD, molecular translocations experiments were performed at 200 mV with this same amplifier by introducing the 2kb DNA at 5 nM in the same 3.6 M LiCl + 10 mM HEPES pH 8 mentioned previously. Results of the DNA experiments were subsequently analyzed using the Nanolyzer (Northern Nanopore Instruments) software.

# Section 3: Results and Discussion

## A. IBD SiN$_X$ Film Analysis

The properties of the 20 nm IBD non-annealed SiN$_x$ films were characterized by using ellipsometry, X-ray reflectometry (XRR), laser-based stress measurement, and wet etch rate tests to study the thickness, density, stress, and wet etch rate (WER), respectively (Figure 3). These measurements, coupled with the nanopore results themselves, are crucial in helping us understand which metrics are important for obtaining a SiN$_X$ film that is suitable for ssNP work, and will help inform the optimization of future depositions.

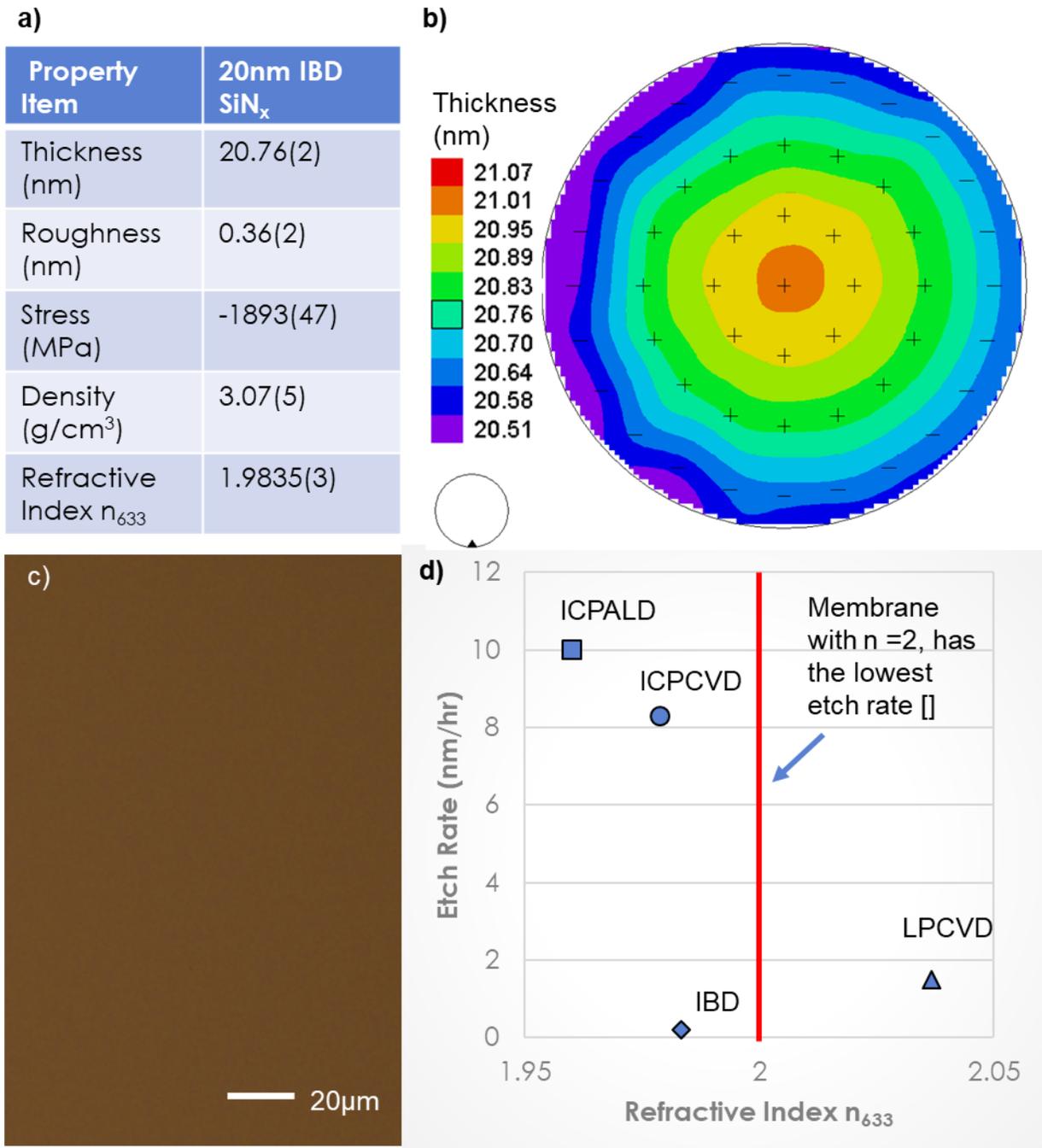

Figure 3. Characterization of 20 nm thick, non-annealed, SiN$_X$ membranes deposited by ion beam deposition (IBD). a) Film properties including thickness, roughness, stress, density, and refractive index of a 20nm SiNx membranes deposited on a silicon wafer by ion beam deposition technique. b) Ellipsometry result showing the wafer scale thickness and uniformity of the film deposited on a six-inch wafer. c) Wet etch test are conducted on the film in hot (80°C) 30 wt.% potassium hydroxide (KOH) bath for 2 hours. The SiNx membranes deposited by IBD techniques remain smooth and continuous after etching test. No obvious defects are observed. d) Plot shows the etch rate and refractive index (n663, refractive index at wavelength of 633nm) comparison of IBD SiNx membrane with membranes deposited by other techniques, low pressure chemical vapor deposition

(LPCVD), inductively coupled plasma chemical vapor deposition (ICPCVD), and inductively coupled plasma atomic layer deposition (ICPALD). the four different membranes.

The ellipsometry results show the IBD film tested had an average thickness of 20.8 ± 0.1 nm with a surface roughness of 0.36 nm and percentage of non-uniformity of 1.2% (Figure 3a,b). In comparison, traditional 20 nm films deposited via LPCVD have a thickness of 19.6 nm with a uniformity of ± 0.2%, as per a recent deposition. While the uniformity of the IBD film is slightly less than that of the LPCVD, this difference represents such a small variation in thickness that the nanopore fabrication and sensing will be unaffected. Thus, in terms of absolute thickness, the IBD films seem to be on par with those deposited by LPCVD.

In contrast, the stress of the $SiN_X$ film is quite different; while LPCVD generates membranes with tensile stress that optically look flat, the nature of the IBD method generates a compressive stress within the $SiN_X$ films, measured to be –1893.6 MPa for our deposition. As a result, the membranes look heavily rippled (Figure 2m). The density of the films also presented a difference, with the $SiN_X$ film measured at 3.07 g/cm$^3$, 96% of the density of stoichiometric LPCVD $SiN_x$ films, 3.2g/cm$^3$ [22]. This measurement suggests that the IBD $SiN_x$ membrane is Si-rich, which could have an impact during the nanopore fabrication as the conductive silicon may provide increased leakage current and/or a path for fabrication to occur. It should be noted, however, that the IBD method allows for the tuning of the Si and N ratio within the thin film, which could be an avenue for optimization in the future.

Lastly, we conducted a KOH wet-etch test on the $SiN_X$ films deposited by IBD and LPCVD $SiN_X$ as well as ICPCVD and PEALD, the two other BOEL-compatible methods previously identified, by subjecting each film to a 2-hour KOH bath at 75°C. While the final workflow for generating BEOL-integrated solid-state nanopore devices may use a membrane release technique other than KOH, the wet etch process can nonetheless reveal the chemical stability of the membrane, which is pivotal for device fabrication and characterization.

After etching, the SiNx films were examined under an optical microscope. The films deposited by IBD and LPCVD appeared smooth and uniform without obvious defects (Figure 3c and Figure S1a), indicating high quality and good chemical stability. However, high-density square defects were observed on the ICPCVD $SiN_X$ film (Figure S1b), where the ICPCVD SiNx had been etched through, partially affecting the Si substrate. Additionally, the presence of scattered defects indicates non-uniform film quality, which could pose challenges for nanopore device fabrication and performance. The PEALD fared even worse, as the $SiN_X$ film was fully removed during the 2-hour wet etch test, leaving only a rough surface of Si wafer (Figure S1c). The rate of the etch was also determined for each deposition method via ellipsometry. The WER of the IBD film was the lowest, at 0.172 nm/hr, with the LPCVD film nearly an order of magnitude greater at 1.475 nm/hr (Figure 3d). The ICPCVD $SiN_X$ displayed a relatively high and non-uniform etch rate, ranging from 8.383 nm/hr to above 10.0 nm/hr and the PEALD $SiN_X$, having been etched all the way through, could only be assigned a minimum value of 10 nm/hr. The high WER observed in PEALD and ICPCVD $SiN_X$ membranes stems from impurity incorporation in the films.

From these analyses, we conclude that of the $SiN_X$ deposition methods that have a low thermal budget, the accuracy and precision of the deposition thickness, the tunability of the Si content, and the high purity of the film all combine to make the IBD $SiN_X$ the best candidate for BEOL-integrated ssNPs.

## B. Solid-State Nanopores

Once mounted in fluidic solution, both IBD non-annealed and IBD annealed membranes were easy to wet and proved to be insulating, as evidenced by the resistance across the chip measuring >1 GΩ. Once the resistance was validated, the membranes underwent fabrication via CBD (Figure 4a,b). As shown, in both types of membranes the voltage ramp generates the expected and characteristic leakage current response prior to the formation of the nanopore (i.e., the breakdown event). While the magnitude of the leakage current is slightly different between the non-annealed and the annealed IBD chips, it is not possible to conclude that the type of film deposition is the cause of this difference, as several variables can contribute to this current including the resistivity of the particular wafer, the thickness of the $SiN_X$ layer, the area of chip exposed to liquid, etc. Additionally, the fabrication voltage is significantly different, but this too depends on extraneous variables, principally the thickness of the thinned region, which itself is dependent on the RIE which has considerable margins of error. As such, we do not draw direct comparisons regarding the breakdown voltage or leakage current, but rather that both exhibit typical pre-fabrication curves which culminate in a sudden increase in current, indicating the transition from leakage and capacitive current to ionic current, and thus a nanopore.

Figure 4: Solid-state nanopore fabrication by controlled breakdown and enlargment by voltage-pulse method on nominally 20nm thick annealed and non-annealed $SiN_X$ membranes formed by IBD. a, b) Representative example of the formation of a solid-state nanopore for non-annealed and annealed IBD membranes, respectively. The nanopores were submerged in 1M KCl, pH 8 solution and the voltage increased at a rate of –3V/min until –5V, and –0.75V/min thereafter until nanopore formation. The red line represents the current, the black dashed line the threshold the software uses to detect nanopore formation, and the blue line the applied voltage. c, d) Growth behaviours observed during the enlargement phase for non-annealed and annealed $SiN_x$ membranes deposited by IBD. Squares represent nanopores that grew fast while triangles represnt nanopores that grew more slowly. Each nanopore was grown in a 3.6 M LiCl pH 8 solution with a square voltage pulse of ± 4V for 3s. e) Typical I-V curves for annealed and non-annealed $SiN_X$ membranes formed by IBD after conditioning, with colours conserved from c,d). f) Distribution of post fabrication pore sizes for annealed and non-annealed $SiN_X$ membranes formed by IBD.

After fabrication, the nanopores were enlarged to their target size of 8 nm via the

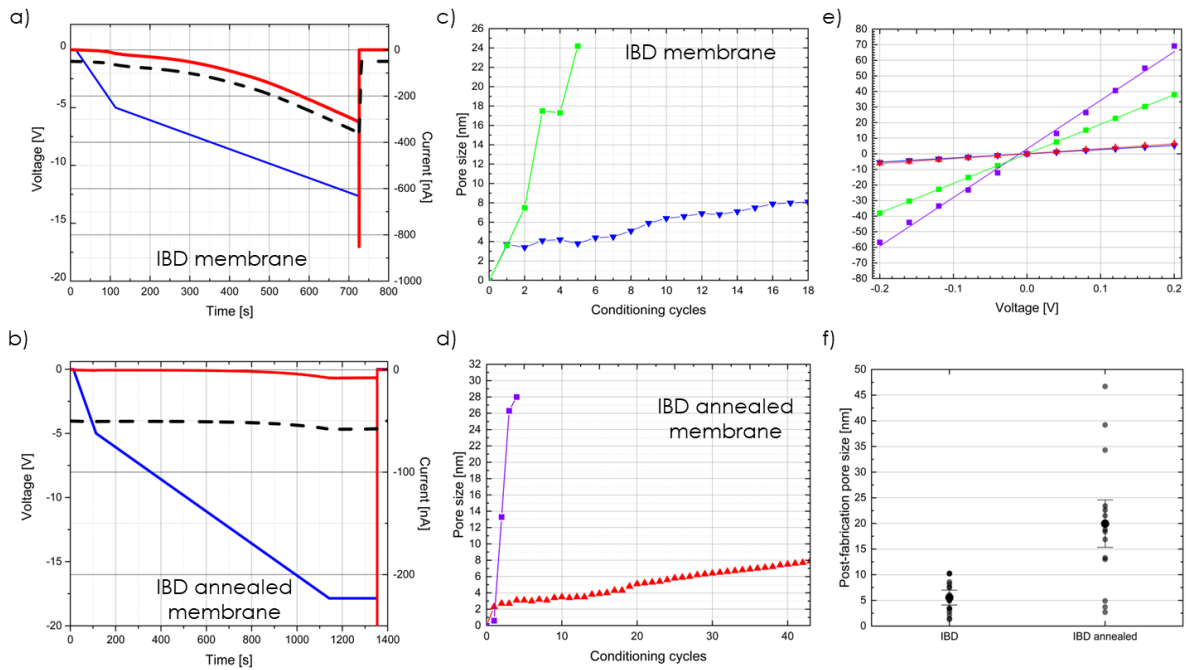

conditioning protocol, and as shown in Figure 4,c,d, nanopores in both non-annealed IBD and annealed IBD films were capable of controlled enlargement (triangles). However, both membrane types also generated pores that were significantly larger immediately after fabrication

(squares). In this case, the target size was set to 20 nm and as shown, the conditioning steps enlarged the pores much more rapidly. It should be noted that this catastrophic pore fabrication and enlargement behaviour is exceedingly rare in typical LPCVD membranes.

Post enlargement, all nanopores underwent a voltage sweep while the current was measured. Characteristic linear I-V curves are shown in Figure 4e, for each pore shown in Figure 4 c,d, demonstrating that pores in both non-annealed IBD and annealed IBD membranes can behave ohmically, and that the rapid pore growth does not preclude linear I-V curves.

In total, 19 nanopores were fabricated on non-annealed IBD chips and 14 on annealed IBD chips. While all these samples underwent successful nanopore fabrication, the distribution of the initial pore sizes of the non-annealed IBD chips was 5.5±3 nm while the distribution of the initial pore sizes for the annealed IBD chips was 20±13 nm (Figure 4f). This large difference in initial pore size was unexpected and may be due to how the annealing modifies the previously amorphous $SiN_X$ structure into one with grains. If CBD causes the breakdown to occur through one of these grains, the resulting pore may be significantly larger. This large initial pore size and rapid enlargement could also prove to be beneficial for applications requiring large (i.e. >20 nm) nanopores as it would obviate the need for potentially lengthy conditioning steps.

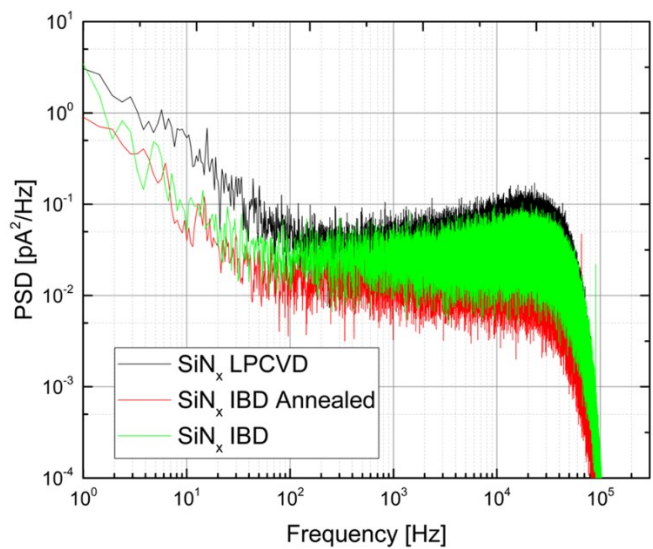

Figure 5: Power spectral density (PSD) of the current trace at the sensing voltage (±200 mV) in 3.6 M LiCl pH 8 for the nanopore devices fabricated on 3 types of membrane devices (IBD, IBD annealed, and LPCVD) showcasing 1/f (or "flicker") noise.

To demonstrate the ability of these nanopores to perform single-molecule measurements, representative PSDs (performed in 3.6M LiCl + 10 mM HEPES (pH 8) and at 200 mV) are presented in Figure 5. For benchmarking, a PSD of a nanopore fabricated in a traditional LPCVD low-stress, silicon-rich, $SiN_x$ membrane is included. Variability in the magnitude of low-frequency 1/$f$ noise is indicative of a stable pore conductance over time (in the seconds timescale), which is generally a good indicator for the viability of the pore for molecular sensing [23]. Here, the nanopores from non-annealed IBD and annealed IBD membranes exhibit a noise profile in the frequency range of 1 Hz – 100 kHz similar to that of nanopores formed in traditional LPCVD.

Finally, single-molecule translocation data using nanopores formed in both non-annealed IBD and annealed IBD membranes is shown in Figure 6. A 15s trace of a 24 nm non-annealed IBD and 8 nm diameter annealed IBD nanopore is shown in Figure 6a and b respectively, where the flow cells contain 3.6M LiCl + 10 mM HEPES (pH 8) and 5 nM of 2 kbp DNA (NoLimits 2000 bp, ThermoFisher). The signal is obtained by applying 200 mV at a sampling rate of 500 kHz before being digitally filtered with a 8-pole Bessel at 100 kHz. Examples of individual events are presented in Figures 6c,d respectively, and the resultant maximum blockage versus dwell time plots are shown in Figures e,f, respectively, validating successful translocation of both single-file and folded DNA molecules. The observed discrepancy between the blockage depths between the non-annealed IBD and annealed IBD nanopores is attributed to a difference in membrane thickness, which could have arisen either during the initial film deposition, or as a result of the local thinning. By using the blockage depths presented for the single-file DNA population, we can extract an effective cylindrical pore thickness by using the following first order approximation: $L = V * \sigma * \pi * \frac{d_{DNA}^2}{4*\Delta I}$, where V is the applied voltage across the nanopore, $\Delta I$ is the dsDNA current blockage and σ is the salt conductivity. This leads to a calculated effective thickness of 25 nm and 14 nm for non-annealed IBD and annealed IBD membranes, respectively, suggesting future work may need to be done to improve the accuracy of the membrane fabrication workflow as both methods deviated from the target thickness of 20 nm by ~25%.

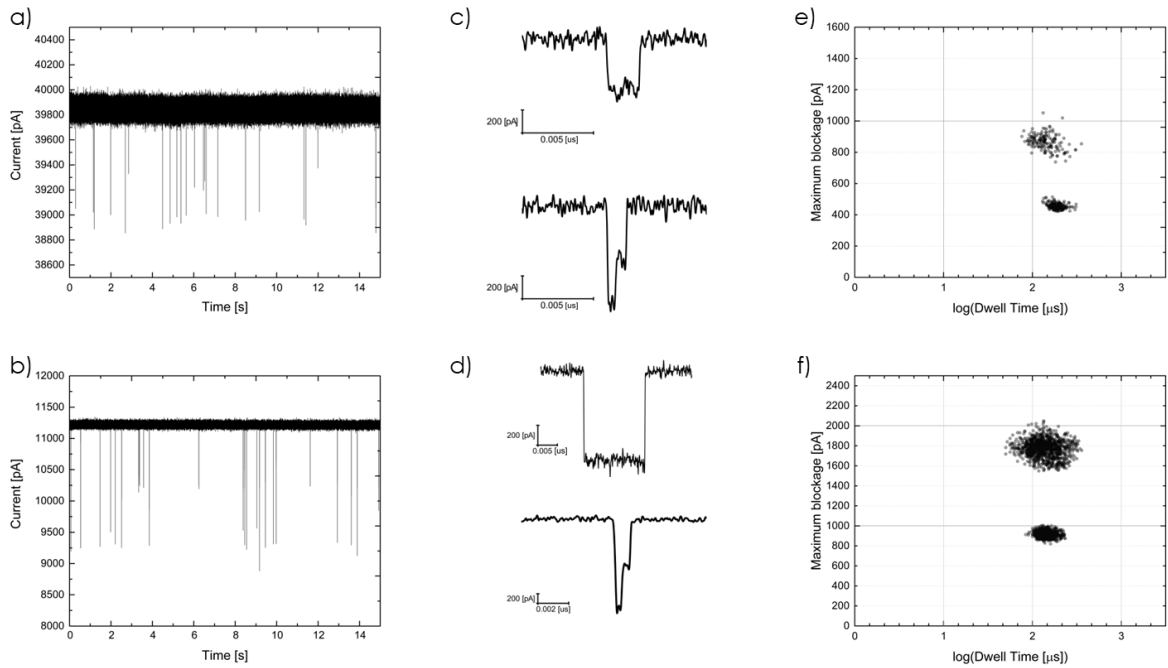

Figure 6: DNA translocation experiments (5 nM 2 kbp no-limit DNA in 3.6 M LiCl in pH 8 at -200 mV, low-pass filtered at 100 kHz) using annealed and non-annealed $SiN_X$ membranes formed by IBD. a,b) Baseline current trace showing several molecular translocations using non-annealed and annealed $SiN_x$ membranes formed by IBD, respectively. c,d) Current traces of typical DNA translocation events observed in non-annealed and annealed $SiN_x$ membranes formed by IBD, respectively. e,f) Scatter plots of a typical maximum blockage current versus dwell time for non-annealed and annealed $SiN_x$ membranes formed by IBD, respectively.

# Conclusion

Our investigation focused on evaluating the quality of $SiN_x$ membranes deposited by ion beam deposition (IBD) onto Si substrates and their suitability for solid-state nanopore biosensing. We showed $SiN_x$ films prepared by IBD show high density, good uniformity, and wet etch resistance, and that low 1/f noise nanopores can be successfully fabricated by controlled breakdown on both annealed and non-annealed IBD $SiN_x$ membranes. These resulting nanopores were then used for single-molecule experiments where 2kbp DNA molecules were translocated and showed a signal-to-noise ratio comparable to nanopore formed in the gold standard LPCVD $SiN_x$ membranes. These findings are promising for the integration of solid-state nanopore systems on CMOS platforms as this represents a Back-End-Of-Line (BEOL)-compatible method of merging solid-state nanopores with CMOS technologies which could significantly advance the development of DNA sensing and storage technologies.

**Acknowledgements:**

This work was supported by Mitacs through the Mitacs Accelerate program. We would also like to thank Vanmaly Phanousith and Simon Riberolles for support with the XRR measurements.

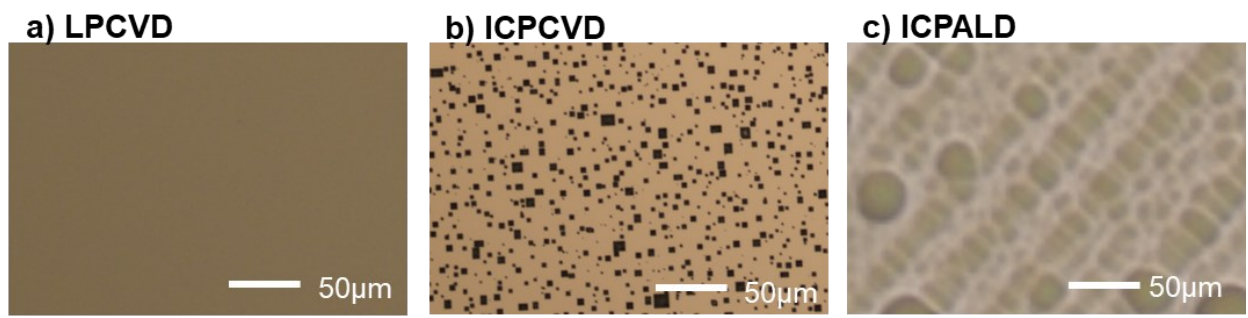

Figure S1, Evaluation of the membrane quality. 20nm silicon nitride (SiNx) membranes are deposited on both sides of silicon wafers by different techniques. The samples are further etched in hot (80°C) 30 wt.% potassium hydroxide (KOH) bath for 2 hours. a,) The SiNx membranes deposited by low pressure chemical vapor deposition (LPCVD) remain smooth and continuous after etching test. No obvious defects are observed. b) The SiNx membrane deposited by inductively coupled plasma chemical vapor deposition (ICPCVD) shows a high density of defects with different sizes across the sample. The SiNx membrane at these defect sites are fully removed. c) The SiNx membrane deposited by inductively coupled plasma atomic layer deposition (ICPALD) is fully etched after the 2-hour test. The image shows the rough etched surface of the underneath silicon.